\documentclass[twocolumn]{aastex62}

\begin{document}


\title{Mutual neutralization in L\MakeLowercase{i}$^+$ -- D$^-$ collisions: a combined experimental and theoretical study}%

\author{Thibaut Launoy}
\affiliation{Institute of Condensed Matter and Nanosciences, Universit\'e catholique de Louvain, 2 Chemin du Cyclotron, 1348 Louvain-la-Neuve, Belgium}
\affiliation{Laboratoire de Chimie Quantique et Photophysique, Universit\'e libre de Bruxelles, 50 av. F.D. Roosevelt, CP160/09, 1050 Brussels, Belgium}
\author{J\'er\^ome Loreau }
\affiliation{Institute of Condensed Matter and Nanosciences, Universit\'e catholique de Louvain, 2 Chemin du Cyclotron, 1348 Louvain-la-Neuve, Belgium}
\affiliation{Laboratoire de Chimie Quantique et Photophysique, Universit\'e libre de Bruxelles, 50 av. F.D. Roosevelt, CP160/09, 1050 Brussels, Belgium}
\author{Arnaud Dochain}
\affiliation{Institute of Condensed Matter and Nanosciences, Universit\'e catholique de Louvain, 2 Chemin du Cyclotron, 1348 Louvain-la-Neuve, Belgium}
\author{Jacques Li\'evin }	
\affiliation{Laboratoire de Chimie Quantique et Photophysique, Universit\'e libre de Bruxelles, 50 av. F.D. Roosevelt, CP160/09, 1050 Brussels, Belgium}
\author{Nathalie Vaeck}
\affiliation{Laboratoire de Chimie Quantique et Photophysique, Universit\'e libre de Bruxelles, 50 av. F.D. Roosevelt, CP160/09, 1050 Brussels, Belgium}
\author{Xavier Urbain }
\email{xavier.urbain@uclouvain.be}
\affiliation{Institute of Condensed Matter and Nanosciences, Universit\'e catholique de Louvain, 2 Chemin du Cyclotron, 1348 Louvain-la-Neuve, Belgium}


\begin{abstract}

We present a combined experimental and theoretical study of the mutual neutralization process in collisions of lithium ions (Li$^+$) with deuterium anions (D$^-$) at collision energies below 1 eV. We employ a merged-beam apparatus to determine total and state-to-state mutual neutralization cross sections. We perform nuclear dynamics calculations using the multi-channel Landau-Zener model based on accurate \textit{ab initio} molecular data. We obtain an excellent agreement between the experimental and theoretical results over the energy range covered in this work. 
We show that the basis sets used in the \textit{ab initio} calculations have a limited influence on the total cross section, but strongly impacts the results obtained for the partial cross sections or the reaction branching ratios. This demonstrates the important role of high-precision measurements to validate the theoretical approaches used to study gas-phase reactive processes. 
Finally, we compute mutual neutralization rate coefficients for Li$^+$ + H$^-$ and Li$^+$ + D$^-$, and discuss their significance for astrochemistry models.

\end{abstract}



\section{\label{sec:intro}Introduction}

The analysis of the chemical composition of astronomical objects is a central problem in astrophysics.
The accurate determination of stellar abundances, in particular, provides insights into stellar and galactic evolution, as well as Big Bang nucleosynthesis. In order to derive these abundances, the use of non-Local Thermodynamical Equilibrium (non-LTE) models is required as departure from LTE is extremely common in these environments \citep{Asplund_2005,barklem_accurate_2016}. A typical issue in non-LTE models arises from uncertainties in reactive or inelastic collisional rate coefficients (or cross sections) involving hydrogen atoms. Among these processes, mutual neutralization (MN) in ion-pair collisions plays an important role in thermalizing atoms owing to the large corresponding cross sections \citep{belyaev_cross_2003,Barklem_2003_LiH}. In MN reactions, oppositely-charged ions collide, resulting in the formation of neutral fragments following electron transfer from the anion to the cation. Mutual neutralization in collisions between atomic or molecular species also plays a crucial role in laboratory plasma physics, photon-dominated regions, planetary ionospheres or in the chemistry of the early universe (see \cite{larsson_ion_2012,Geppert_2013,Hedberg_2014} and references therein). 

From a more fundamental point of view, ion pair states can be seen as an analog to Rydberg states, in which the electron is replaced by the anion \citep{reinhold_2005}. These long-range excited states manifest themselves in the time-resolved photodissociation of bi-alkali molecules \citep{rosker_1988} via the long recurrence times of the vibrational wavepackets. Such long-range vibrational states are aptly described in the multichannel quantum defect theory (MQDT) framework, which accounts for the influence of the covalent states on the lifetime of individual levels and the appearance of local perturbations in the spectra \citep{vieitez_2009}. Heavy Rydberg states have interesting properties such as a large dipole moment, and can be employed to determine accurate electron affinities and bond dissociation energies \citep{mollet_2010,meyer_2018}. Moreover, the formation of these heavy Rydberg states has been suggested as a way to realize cold, strongly-coupled plasmas \citep{kirrander_2013}.

Recently, significant progress has been accomplished in the theoretical description of mutual neutralization both with rigorous quantum calculations \citep{croft_theoretical_1999,guitou_mg-h_2012,Hedberg_2014,larson_theoretical_2016,mitrushchenkov_calcium-hydrogen_2017} or using various asymptotic models \citep{yakovleva_atomic_2016,Barklem_2017,barklem_excitation_2016,
belyaev_theoretical_2014,deruette_2018_prl}, in various collision energy ranges. This includes the theoretical study of mutual neutralization between H$^-$ and several elements such as H$^+$ \citep{stenrup_mutual_2009,nkambule_differential_2016}, He$^+$ \citep{larson_theoretical_2016}, Li$^+$ \citep{croft_rate_1999,croft_theoretical_1999,belyaev_cross_2003,Belyaev_2018_LiH}, Be$^+$ \citep{yakovleva_atomic_2016,Hedberg_2014}, O$^+$ \citep{Barklem_2017}, Na$^+$ \citep{Dickinson_NaH_1999}, Mg$^+$ \citep{belyaev_cross_2012,guitou_mg-h_2012}, Al$^+$ \citep{belyaev_inelastic_2013}, Si$^+$ \citep{Belyaev_2014_SiH}, Ca$^+$ \citep{barklem_excitation_2016, mitrushchenkov_calcium-hydrogen_2017} or Cs$^+$ \citep{belyaev_theoretical_2014}.

Despite this, the experimental investigation of this kind of system remains an important challenge. Earlier experimental MN studies between atomic species \citep{Olson_1970,peart_new_1985,peart_measurements_1987,Peart_1989_NO_OO,
Peart_1992,peart_merged_1994,nkambule_differential_2016} have allowed measurements of cross sections, but few of them have succeeded in completely characterizing these systems by providing both cross sections and branching ratios of the neutral products formed. Moreover, these experiments  have mostly been performed at collision energies above 1 eV although the mutual neutralization process is also crucial at lower energies. The first complete subthermal study of quantum-state-resolved MN processes was recently performed by \citet{deruette_2018_prl} for the O$^+$ + O$^-$ and N$^+$ + O$^-$ systems. 
In this context, even one of the simplest systems where MN plays a role, i.e. the mutual neutralization reaction between lithium and hydrogen ions, is an experimental challenge. 

Although its abundance is low, lithium has a singular significance in astrophysics (see \cite{stancil_lithium_1996,Barklem_2003_LiH, Asplund_2005} and references therein). Its abundance is a key parameter for stellar atmospheres, models of Big Bang nucleosynthesis, and the chemistry of the early universe. Collisions of lithium with atomic hydrogen and deuterium are important for the thermalization of Li and for non-LTE modelling via several processes such as (de-)excitation, ion-pair production (Li $+$ H $\rightarrow $ Li$^+ +$ H$^-$) and mutual neutralization \citep{Barklem_2003_LiH,Stancil1998}. Therefore, accurate theoretical and experimental data on these processes are needed in order to validate theoretical models.

In this work, we investigate the MN reaction (\ref{eq:MN_LiH}) both theoretically and experimentally, at energies below 1 eV: 
\begin{equation}
\textrm{Li}^+ (1s^2) + \textrm{H}^-/ \textrm{D}^- \rightarrow \textrm{Li}^* (1s^2\, nl) +  \textrm{H}/ \textrm{D} \;(1s)
\label{eq:MN_LiH}
\end{equation} 
Few experimental studies have been conducted about reaction (\ref{eq:MN_LiH}). Using an inclined-beam set-up, absolute cross sections have been measured by \citet{peart_measurements_1987} for collision energies above 33 eV. Several years later, \citet{peart_merged_1994} employed a merged beam experiment to go down to 0.7 eV of collision energy. However, these two studies did not provide absolute cross sections at subthermal energies nor the branching ratios of the neutral products. 
Several theoretical studies were performed on reaction (\ref{eq:MN_LiH}). Early calculations were realized by \citet{Bates_1956} within the framework of the Landau-Zener (LZ) method. Using an asymptotic model to estimate the non-adiabatic couplings and with the LZ model, \citet{janev_ion-ion_1978} refined these results for the alkali atoms. It was shown that at energies below 1 eV, the dominant contribution to the cross section arises from capture into the $(n+1)s$ atomic state, where $n$ is the principal quantum number of the valence electron of the alkali atom in the ground state. On the other hand, at collision energies above 100 eV, the first ($np$) excited atomic state contributes significantly to the total cross section. 
This has been confirmed by several subsequent studies. For instance, \citet{mendez_molecular_1990} calculated cross sections at energies above 25 eV based on \textit{ab initio} potential energy curves and non-adiabatic couplings, while \citet{ermolaev_mutual_1992} computed partial and total cross sections for energies above 100 eV by means of a one-active electron model and a large atomic basis set. With a similar procedure initially employed at a collision energy of 375 eV, \citet{Lin_1996} extrapolated their results to lower energies assuming the same transition probabilities. The agreement between their results and experimental data is rather good at 0.7 eV and above 50 eV, but between these energies, the total cross section is considerably underestimated.
\citet{croft_theoretical_1999} performed non-adiabatic quantum nuclear dynamics calculations to provide partial and total cross sections at energies ranging from 1 meV to 10 eV for the Li$^+$ + H$^-$ MN reaction \citep{croft_rate_1999} and from 0.68 to 40.1 eV for the Li$^+$ + D$^-$ system. Their total cross sections were in good agreement with experimental results. 
Based on the same \textit{ab initio} data, \citet{Belyaev_2018_LiH} recently refined these results with the use of the quantum hopping probability current method and provided mutual neutralization rate coefficients for various temperatures that fall in agreement with the previous results of \citet{croft_theoretical_1999}. 

The paper is organised as follows. The experimental set-up is briefly presented in Sec. \ref{sec:exp}. Computational details of quantum chemistry  and nuclear dynamics calculations are described in Sec. \ref{sec:theory}. We discuss our experimental and theoretical total and partial cross sections in Sec. \ref{sec:results}. These results are subsequently employed to determine the rate coefficients for reaction (\ref{eq:MN_LiH}).

\section{\label{sec:exp}Merged beam setup} 

\begin{figure*}
\begin{center}
\includegraphics[width=.8\textwidth]{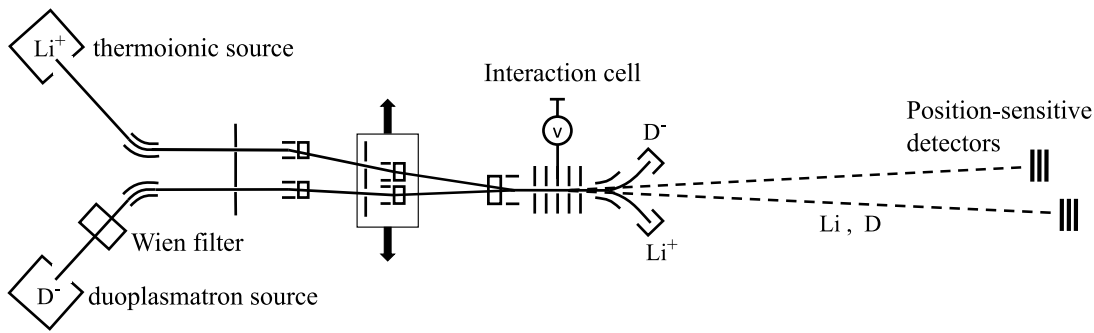}
\caption{Experimental set-up used for the mutual neutralization process (not to scale).}
\label{fig_exp_setup}
\end{center}
\end{figure*}

The mutual neutralization measurements were performed using the merged beams setup at the Universit\'e catholique de Louvain, depicted schematically in Fig. \ref{fig_exp_setup}. This experimental setup has already been described by \cite{deruette_2018_prl} (MN detection system), \cite{Olamba1996} (merged beam improvement), and \cite{Cherkani1991} (merged beam system).

The beam of negative ions D$^-$ is produced from a duoplasmatron source filled with D$_2$, and a Wien filter is employed to select the ion masses. The beam of positive ions $^7$Li$^+$ is produced using an isotope-enriched thermoionic emitter mounted in Pierce extraction geometry. As this source only emits $^7$Li$^+$ it does not require a mass filter element. The cations and anions beams are accelerated at 13500 eV and 4500 eV respectively, and are shaped by ion optics and collimators. They are then merged in an electrically-biased interaction cell of adjustable length from 2 to 7.6 cm where the reaction occurs. The collision energy can be tuned by varying the voltage on this cell, allowing the measurement of mutual neutralization cross sections from about 10 eV down to the meV range. The lower limit is set by the angular distribution of the beams and their slight misalignement (below 0.5 mrad).

After the interaction the beams are separated and collected using polarized Faraday cups giving $I^+$ and $I^-$ currents. The neutral reaction products are not deflected and fly directly to a three-dimensional imaging detection system. It consists of two time- and position-sensitive detectors triggered in coincidence. Both are Z-stacks of 4~cm diameter microchannel plates (MCP) placed in front of a resistive anode. The dead region between them is reduced to 2 cm by mounting them 10 cm apart along the beam axis. In order to ensure that the beams are aligned, we photodetach the D$^-$ anions in the observation cell with a CW diode laser and check that the neutral products hit the edge of one of the MCPs. Then, once  the laser is turned off, we adjust the Li$^+$ beam and the mixing deflectors in order to maximize the coincidence rate. As a result, the center of mass trajectory is not centered between the detectors. This increases the coincidence rate due to the fact that the recoil velocity of $^7$Li$^\ast$ is smaller than the recoil of D, causing the laboratory frame trajectories of Li to depart less from the beam axis than those of D atoms.
The detection system can only detect a fraction of the solid angle for a given recoil energy (the kinetic energy release, or KER). 
As the collision energy is controlled by the angular distribution of the Li$^+$ and D$^-$ beams at matched velocities, the orientation of the collision axis is randomized, and the recoil angular distribution is also uniform. It is then easy to rebuild an acceptance map for the mutual neutralization reaction.

The use of two separate detectors allows for the simultaneous detection of the two products, and the long drift distance of 3.25 m from the interaction cell to the imaging detectors allows us to minimize the misalignment of the two beams. This is done by optimizing the MN coincidence rate relying on the (expected) $1/E_{\text{CM}}$ energy dependence of the cross section and the fact that the angular dispersion of the beams is  the main limiting factor for the resolution in the definition of the center-of-mass collision energy, $E_{\text{CM}}$. Considering a collision between an anion of mass $m_A$ and kinetic energy $E_A$, and a cation of mass $m_B$ and kinetic energy $E_B$ ($E_A$ and $E_B$ are the ion-beam energies in the laboratory system), we get
\begin{displaymath}
E_{\text{CM}} = \mu \left( \frac{E_A}{m_A} +  \frac{E_B}{m_B}  - 2 \sqrt{\frac{E_A E_B}{m_A m_B} }\cos{\phi} \right) 
\end{displaymath}
where $\mu =m_A m_B /(m_A + m_B)$ is the reduced mass and $\phi$ is the angle between the ion trajectories. In the limit of matched velocities, \textit{i.e.} $E_A/m_A=E_B/m_B$, the above equation reduces to $E_{\text{CM}} \simeq \frac{\mu}{m_A}E_A \phi^2$.

The ratio between the finite length of the interaction region and the distance to the detectors determines the KER resolution. A longer flight distance gives a better precision on the velocity measurements of the neutrals, and thus it increases the resolution but also limits the angular acceptance. Here, we reach a resolution of 10 meV FWHM at 1 eV of KER, and we are thus able to separate the contributions of the electronic states of the neutral reaction products.

\section{\label{sec:theory}Theoretical methods}
\subsection{Electronic structure calculations}

In mutual neutralization reactions, the reactants approach along a $1/R$ Coulomb potential. Therefore, the main difficulty in treating MN reactions is to describe correctly all the molecular electronic states below the ion-pair limit over a wide range of internuclear distances. Table~\ref{table_LiH} summarizes the Li$(1s^2\, nl$) + H($1s$) asymptotic dissociation channels below the ion-pair limit, the molecular states of LiH that emerge from these dissociation limits, and the internuclear distance $R_x$ at which the potential energy curves (PECs) for the ion-pair and covalent channels are expected to cross based on the atomic energies.

\begin{table*}
\begin{center}
\begin{tabular}{cccc}
\hline
Asymptotic & Molecular	& Experimental   &  $R_x$ ($a_0$)\\
atomic states & states & atomic energies (eV)  & \\
\hline
\hline
Li($1s^2 2s$)   + H($1s$) 		& $^{1,3}[\Sigma^+]$ 				& 0		& 7.413	\\
Li($1s^2 2p$)   + H($1s$) 		& $^{1,3}[\Sigma^+, \Pi]$  				& 1.848	& 10.61	\\
Li($1s^2 3s$)   + H($1s$) 		& $^{1,3}[\Sigma^+]$   				& 3.373	& 21.64	\\
Li($1s^2 3p$)   + H($1s$)  		& $^{1,3}[\Sigma^+, \Pi]$  			& 3.834	& 33.70	\\
Li($1s^2 3d$)   + H($1s$)  		& $^{1,3}[\Sigma^+, \Pi, \Delta]$	& 3.879	& 35.63	\\
Li($1s^2 4s$)   + H($1s$)  		& $^{1,3}[\Sigma^+]$   			& 4.341	& 89.78	\\
Li($1s^2 4p$)   + H($1s$)   		& $^{1,3}[\Sigma^+, \Pi]$ 			& 4.522	& 222.3	\\
Li($1s^2 4d$)  + H($1s$)  		& $^{1,3}[\Sigma^+, \Pi, \Delta]$		& 4.541	& 263.3	\\
Li($1s^2 4f$)   + H($1s$) 		& $^{1,3}[\Sigma^+, \Pi, \Delta, \Phi]$	& 4.542	& 265.5	\\ 
Li$^+$ ($1s^2$) + H$^-$(1$s^2$) 	& $^1\Sigma^+$  				& 4.639	&		\\
\hline
\hline
\end{tabular}
\end{center}
\caption{Asymptotic dissociation channels correlated to the LiH molecular states lying below the ion-pair limit. The atomic energies are taken from the NIST database \citep{NIST_ASD}. $R_x$ is the  distance where the Li$^+$ + H$^-$ ion-pair channel is expected to cross each Li + H covalent channel, based on the atomic energies.}
\label{table_LiH} 
\end{table*}

Several basis sets were used to compute the PECs of LiH. In all cases the hydrogen atom was described by the aug-cc-pV5Z basis set augmented by additional optimized functions that accurately reproduce the hydrogen electron affinity \citep{loreau_ab_2010}. Different basis sets were tested for the lithium atom in order to evaluate the influence of the basis set on the dynamical results, \textit{i.e.} total and partial cross sections: an aug-cc-pCV5Z basis set (ACV5Z in the following), an ACV5Z basis set augmented by even-tempered functions located on the lithium atom as in \cite{gim_studies_2014} or fixed in the middle of the lithium-hydrogen bond -- respectively referred to as ET-Li and ET-mid in the following. Using the {\small AUTOSTRUCTURE} package, linear combinations of Slater type orbitals describing the $n=1-4$ $s,p,d$ lithium orbitals were obtained \footnote{P. Quinet and P. Palm\'eri, private communication} and fitted to gaussian-type functions. This basis set will be referred to as ACV5Z+G in the following. We have also used the basis set recently developed by \cite{gim_studies_2014} to describe the excited states of the lithium atom up to $n=6$. However, we were unable to reproduce their results.

Table~\ref{tab_atomic_en} shows the energy of the electronic states of Li using these basis sets calculated with the Multi-Reference Configuration Interaction (MRCI) \citep{knowles_efficient_1988,knowles_internally_1992} program implemented in the {\small MOLPRO} package \citep{MOLPRO_2015}. The results obtained with the ACV5Z+G basis set are in  good agreement with experimental energies whereas the ET-mid basis set is less suitable for the Li($4s$) state, and the ACV5Z basis set incorrectly describes both the $3d$ and $4s$ atomic states. The quality of the basis set will have an important impact on the dynamical results (cross section and branching ratios), as will be shown below.

\begin{table}
\begin{center}
\begin{tabular}{c|ccccc}
\hline
Atomic state		& NIST 	& ACV5Z 	& ACV5Z+G	& ET-mid	\\ \hline \hline
Li($2s$)	 	&0		&0		&0			&0 		\\
Li($2p$)		& 1.8478	& 1.8497	& 1.8495 		& 1.8497 	\\
Li($3s$) 		& 3.3731	& 3.3809	& 3.3721 		& 3.3719 	\\
Li($3p$)		& 3.8342	& 3.8358	& 3.8357 		& 3.8358 	\\
Li($3d$)		& 3.8786	& 4.0425	& 3.8770 		& 3.8772 	\\
Li($4s$) 		& 4.3409	& 4.6921	& 4.3561 		& 4.4006 	\\
\hline
\hline
\end{tabular}
\end{center}
\caption{\label{tab_atomic_en}
Lithium atomic energies (in eV) using various basis sets, compared to the reference NIST values.}
\end{table}

\begin{figure}
\includegraphics[width=0.45\textwidth]{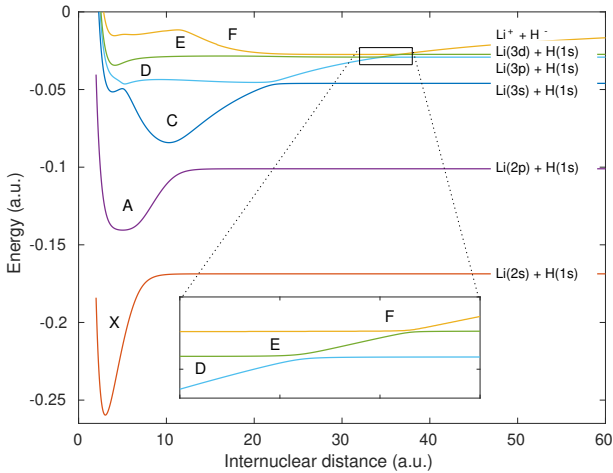}
\caption{Adiabatic potential energy curves of the six lower $^1\Sigma^+$ states of LiH obtained with the ACV5Z+G basis set. The state labels are those introduced by \cite{Boutalib_1992}.}
\label{fig_PEC}
\end{figure}

It is well known that the crossings between the PECs that occur at very large internuclear distances can be considered as diabatic and do not significantly affect the MN reaction. The existence of an optimal window of crossing distance ranging from $\sim10 a_0$ to $\sim40 a_0$ was previously highlighted \citep{belyaev_inelastic_2013}. In consequence, the $4p$, $4d$, and $4f$ electronic states of Li will not be included in the present study. Moreover, as the molecular state dissociating into the ion-pair is a $ ^1\Sigma^+$state, only states of that symmetry were considered.
\textit{Ab initio} molecular structure calculations have been performed for the seven lowest $^1\Sigma^+$ electronic states that take part in the MN reaction (\ref{eq:MN_LiH}) with the {\small MOLPRO} package using the State-Averaged Complete Active Space Self-Consistent Field (SA-CASSCF) approach \citep{knowles_efficient_1985,werner_efficient_1988} followed by internally contracted MRCI computations. The latter calculations include the Davidson correction \citep{langhoff_configuration_1974} to correct for the effect of quadruple excitations. 
The PECs were calculated using every tested basis set on a grid from $R=2 a_0$ to $R=60 a_0$ with a step of 0.1 $a_0$ or less, with an increased density of points close to the avoided crossings. The results for the ACV5Z+G basis set are shown in Figure~\ref{fig_PEC}. The last molecular state goes diabatically to the ion-pair dissociation limit at $R=\infty$. The non-adiabatic coupling matrix elements (NACME) were calculated on the same grid with the three-point method as implemented in {\small MOLPRO}. The PECs are similar to the ones reported in the literature \citep{mendez_molecular_1990,Boutalib_1992,gadea_accurate_2006,gim_studies_2014} and have the same global aspect as the potential energy curves for other alkali hydrides such as NaH or MgH \citep{Dickinson_NaH_1999,guitou_mg-h_2012} with a clear influence of the ion-pair channel.

Based on the NACMEs, the position of the avoided crossings and the maximum amplitude of the couplings obtained with the various basis sets are identified and reported in Table~\ref{table_Rx}.

The results for the inner avoided crossings for the $X-A$ and $A-C$ $^1\Sigma^+$ states are in agreement with the literature (see \cite{mendez_molecular_1990,croft_theoretical_1999}) for every basis tested, with the exception of the ET-mid basis set for which the maximum amplitude of the couplings is larger. Major discrepancies appear for the long-range crossings. In particular, the position of the avoided crossing as well as the value of the coupling for the $D-E$ and the $E-F$ $^1\Sigma^+$ states show a wide variation depending on the basis set used to describe the lithium atom. This will have a strong impact on the cross sections, as detailed below.

\begin{table*}
\begin{center}
\begin{tabular}{l|cc|cc|cc|cc|cc}
Crossing 	& \multicolumn{2}{c}{$X-A$}	& \multicolumn{2}{c}{$A-C$}	& \multicolumn{2}{c}{$C-D$}	& \multicolumn{2}{c}{$D-E$}	& \multicolumn{2}{c}{$E-F$}	\\ \hline 
ACV5Z 		& 6.70 	& 0.20 	& 10.60 	& 0.16 	& 19.60 	& 0.36 	& 28.45	& 0.42 	& 36.60  &  2.10 	\\
ACV5Z+G		& 7.20 	& 0.21 	& 11.30 	& 0.18 	& 22.05 	& 0.50 	& 34.40 	& 1.24 	& 35.90  &  2.04	\\
ET-mid	  	& 6.80 	& 0.48 	& 11.00 	& 0.41 	& 21.40 	& 0.90 	& 32.88 	& 1.81 	& 36.88  & 3.44		\\ 
ET-Li 		& 6.71 	& 0.20 	& 10.64 	& 0.16 	& 19.97 	& 0.37 	& 29.32 	& 0.46 	& 30.08  & 0.62 	\\ 
\hline \hline
\end{tabular}
\end{center}
\caption{\label{table_Rx}
For each of the avoided crossings between the lowest $^1\Sigma^+$ molecular states of LiH (see Fig. \ref{fig_PEC}), the left column is the position of the crossing $R_x$ (in a.u.) and the right column is the value of the non-adiabatic coupling (in a.u.) at the crossing point.}
\end{table*}

\subsection{\label{sec:dynamics} Nuclear dynamics}

The position of the avoided crossings and the non-adiabatic couplings are employed to compute theoretical partial and total cross sections by means of the multi-channel Landau-Zener approach. This method is known to be particularly well suited for processes such as the mutual neutralization reaction in which the attractive PEC of the entrance channel crosses a series of PECs corresponding to the neutral fragments, with results in good agreement with fully quantum-mechanical calculations \citep{yakovleva_atomic_2016,belyaev_cross_2003,Hedberg_2014,Barklem_2017,belyaev_cross_2012,
belyaev_inelastic_2013,Belyaev_2014_SiH,barklem_excitation_2016,mitrushchenkov_calcium-hydrogen_2017,belyaev_theoretical_2014}. 

In the Landau-Zener model, a transition probability is associated to each avoided crossing between the ionic and the covalent PECs (see Fig. \ref{fig_PEC}). 
For the $k$th channel, the transition probability $p_k$ is given by
\begin{equation}
p_k=\exp \left(- \frac{2\pi \, H_{ik}^2 (R)}{v_R \, \Delta F(R)} \right) \bigg|_{R=R_{x,k}}
\label{eq:pn_LZ}
\end{equation}
where $R_{x,k}$ denotes the position of the crossing, $H_{ik}$ is the coupling matrix element (directly related to the non-adiabatic coupling matrix element \citep{Nikitin_1984,Nakamura_1997}), $\Delta F(R)$  is the difference of slopes of the ionic and covalent PECs, and $v_R$ is the radial collision velocity:
\begin{eqnarray}
& v_R & = v_0 \left(1 - \frac{V(R_x)}{E} - \left(\frac{b}{R_x}\right)^2 \right)^{1/2} \\
& & \simeq v_0 \left(1 + \frac{1}{R_xE} - \left(\frac{b}{R_x}\right)^2 \right)^{1/2}
\end{eqnarray}
in which $b$ is the impact parameter, $v_0$ is the initial collision velocity, $\mu$ is the reduced mass of the system, and the energy is $E=\mu v_0^2/2$. The crossings are treated as independent, an approximation that is questionable in the case of the crossings between the ion-pair channel and the Li($3d$) + H and Li($3p$) + H channels which are separated by $2a_0$ only.

In the case of $N$ covalent channels, the total transition probability $P_k$ from the ionic channel to the $k$th exit covalent channel is, when omitting the accumulated phase along the trajectories:
\begin{widetext}
\begin{eqnarray}
P_k  & = & p_1 p_2 \ldots p_k (1-p_k) \Bigg[ 1 + \left( p_{k+1}p_{k+2} \ldots p_N\right)^2  \nonumber \\ \nonumber
	 &&+ \left(1-p_{k+1}\right)^2 + p_{k+1}^2\left(1- p_{k+2}\right)^2 + \left(p_{k+1}p_{k+2}\right)^2 \left(1-p_{k+3}\right)^2 \\ \nonumber
	 &&+ \ldots + \left(p_{k+1}p_{k+2} \ldots p_{N-1}\right)^2 \left(1-p_N\right)^2 \Bigg] \qquad ; \qquad k < N-1 \\\nonumber
P_{N-1} &= & (p_1 p_2 \ldots p_{N-1}) (1-p_{N-1}) \left[ 1 + p_N^2  + (1-p_N)^2 \right]  \\\nonumber
P_N &= & 2p_1 p_2 \ldots p_{N} (1-p_{N}) 
\end{eqnarray}
\end{widetext}
where $p_k$ is given by Equation~(\ref{eq:pn_LZ}), and the channels are numbered in the order of appearance of the avoided crossings, $N$ being the innermost avoided crossing.

The partial cross section for capture into the covalent channel $k$ is then given by 
\begin{displaymath}
\sigma_k = 2\pi \int_0^{b_{x,k}} P_k (b)\, b \, db
\end{displaymath}
where $b_{x,k}$, the largest impact parameter for which the crossing is still accessible, is expressed as 
\begin{displaymath}
b_{x,k} \simeq R_{x,k} \left( 1 + \frac{1}{R_{x,k}E}\right)^{1/2} .
\end{displaymath}

The total mutual neutralization cross section is simply given by the sum of partial cross sections.


\section{Results and discussion}\label{sec:results}
\subsection{Total cross section}

\begin{figure}[h]
\includegraphics[width=0.53\textwidth]{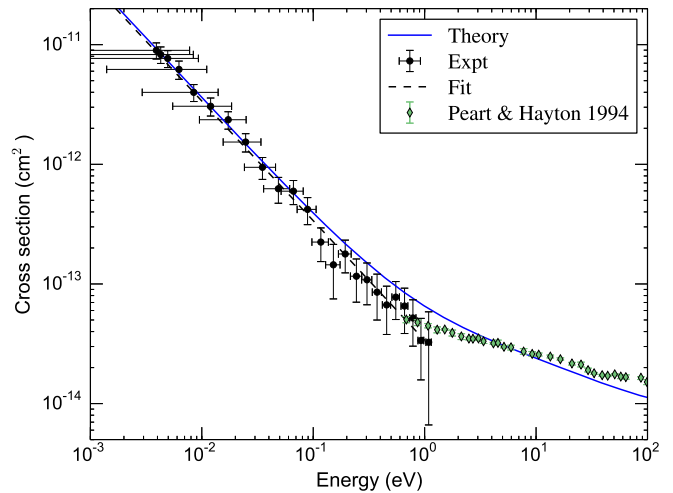}
\caption{Measured total cross section for the mutual neutralization in Li$^+$ + D$^-$ $\rightarrow$ Li$^\ast$ + D (black circles). 
The dashed black line is the $E^{-1}$ cross section that provides the best fit to the experimental data (see text). The solid blue line is the theoretical result using the ACV5Z+G basis set. The diamonds are the measurements of \citet{peart_merged_1994}.}
\label{fig_cs_LiD_exp}
\end{figure}

The measured total mutual neutralization cross section for Li$^+$ + D$^-$ collisions is shown in Fig.~\ref{fig_cs_LiD_exp} as a function of the collision energy from 1.1 eV down to 3.9 meV assuming a 50\% detection efficiency of the MCP and taking into account geometrical corrections. 
The experimental cross section is affected by uncertainties both in collision energy and magnitude. The former originates from the actual relative velocity distribution in the co-moving frame of the beams, which tends to a Maxwellian distribution with $T \simeq 50$~K at matched velocities, while the latter reflects the rapidly dropping statistics with increasing collision energy. Note that all cross section values are affected by systematic uncertainties ($\approx 15$ \%) added in quadrature to the statistical uncertainty given at the 90\% confidence limit.
The cross section agrees well with the previous measurements by \citet{peart_merged_1994} in the overlapping collision energy range (about 1 eV). 
Two regimes can be distinguished: below collision energies of 0.5 eV, the cross section behaves as $E^{-1}$ as expected from the Wigner threshold law \citep{Wigner1948,padellec_investigation_2017} and in agreement with previous mutual neutralization studies \citep{stenrup_mutual_2009,nkambule_differential_2016,Hedberg_2014}, while above 0.5 eV the cross section becomes flatter. This behavior is well reproduced by our calculations based on the multichannel Landau-Zener approach with the molecular data obtained with the ACV5Z+G basis set, shown by the full line in Fig.~\ref{fig_cs_LiD_exp}. 
A fit of the experimental data to the expression $\sigma=aE^{-b}$, shown by the dashed line in Fig.~\ref{fig_cs_LiD_exp}, provides the following values for the parameters: $a=  (3.24 \pm 0.56)\times 10^{-14}$ cm$^2$eV and $b = 1.01 \pm  0.07$, which can be compared to the theoretical value of $a= 3.70 \times 10^{-14}$ cm$^2$eV.  The value of parameter $b$ is fully compatible with the Wigner threshold law for Coulomb processes.

\begin{figure}[h]
\includegraphics[width=0.53\textwidth]{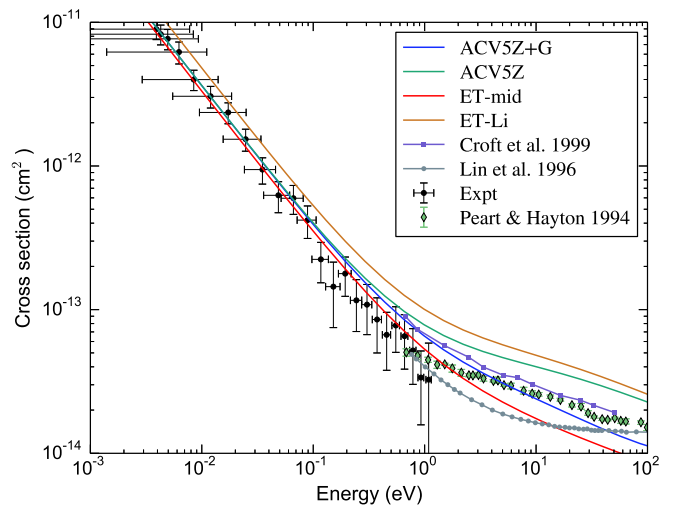}
\caption{Comparison of the theoretical total cross section for Li$^+$ + D$^-$ mutual neutralization obtained with different basis sets (see text for details) and compared to the present experimental results, to the experimental data of \citet{peart_merged_1994}, and to the theoretical calculations of \citet{croft_theoretical_1999} and \citet{Lin_1996}.}
\label{fig_cs_LiD_th}
\end{figure}

At low energy the total cross section is not very sensitive to the basis set used to perform the \textit{ab initio} calculations, as illustrated in Fig. \ref{fig_cs_LiD_th}, although the cross section obtained with the ET-Li basis set is too large. On the other hand, at high energy strong discrepancies between the results obtained with the various basis sets appear. Our measurements and those of \citet{peart_merged_1994} are best reproduced with the ACV5Z+G and ET-mid basis sets. 
At a collision energy of 1 eV the ACV5Z+G and ET-mid cross sections are larger than the experimental results of  \cite{peart_merged_1994} by 50\% and 20\%, respectively, while at 100 eV the ACV5Z+G and ET-mid results are respectively 25\% and 50\% smaller, so that the energy dependence of the cross section does not fully agree with the measurements of \citet{peart_merged_1994}. A similar observation was made in the case of H$^+$--H$^-$ mutual neutralization for energies between 3 and 100 eV \citep{nkambule_differential_2016,Peart_1992}.
The ACV5Z+G cross section also shows a good agreement with the theoretical calculations of \citet{croft_theoretical_1999}. There is on the other hand a  discrepancy with the theoretical results of \citet{Lin_1996} that is probably due to the fact that their cross sections at low energy were obtained by an extrapolating procedure, as discussed by \cite{croft_theoretical_1999}.

\subsection{Partial cross sections and branching ratios}

\begin{figure}[h]
\includegraphics[width=0.5\textwidth]{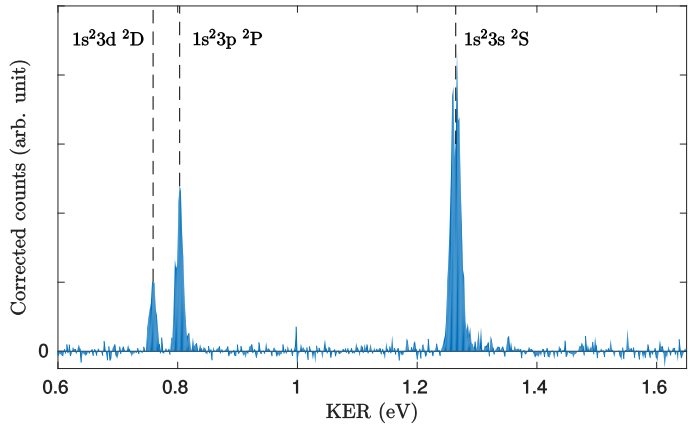}
\caption{KER spectrum for the $^7$Li$^+$ + $^2$D$^-$ $\rightarrow$ Li($1s^2\ nl$) + D($1s$) mutual neutralization reaction measured at a collision energy of 3.9 meV. The dashed lines indicate the position of the Li($n=3$) states.}
\label{fig_ker}
\end{figure}

The experimental KER distribution corresponding to mutual neutralization in Li$^+$ + D$^-$ collisions recorded at 3.9 meV with a 2 cm long interaction cell is shown in Fig. \ref{fig_ker}. Each peak is due to a single excited state of Li, while D is in the ground state following the loss of its extra electron. Due to the resolution of about 10 meV, we can unambiguously assign the peaks to the $3s$, $3p$, and $3d$ states of Li, even though the $3p$ and $3d$ states are separated by only 44 meV.
The analysis of the KER results shows that the Li($n=4$) states do not contribute to the mutual neutralization cross section, as expected based on the large distances at which these electronic states interact with the ion-pair state (see Table \ref{table_LiH}). Moreover, it also demonstrates that in the range of collision energies investigated in this work the lowest electronic states, Li($n=2$), are not populated after the reaction either. 

\begin{figure}[h]
\includegraphics[width=0.5\textwidth]{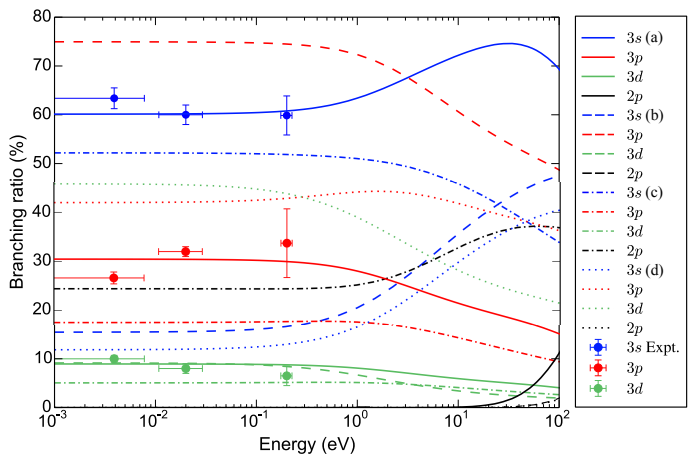}
\caption{Branching ratios for the mutual neutralization reaction Li$^+$ + D$^-$ $\rightarrow$ Li ($1s^2\  nl)$ + D($1s$) measured at three collision energies compared to the calculations based on the \textit{ab initio} data obtained with (a) the ACV5Z+G basis set (full lines), (b) the ACV5Z basis set (dashed lines), (c) the ET-mid basis set (dot-dashed lines), and (d) the ET-Li basis set (dotted lines).}
\label{fig_BR}
\end{figure}

The area of each peak in this KER spectrum gives access to the branching ratios and the partial (state-selective) cross sections. The KER has been analyzed at three collision energies (3.9 meV, 20 meV, and 200 meV), and the resulting experimental branching ratios are shown in Fig. \ref{fig_BR} together with the results obtained from the partial cross sections calculated with the multichannel Landau-Zener model based on the \textit{ab initio} ACV5Z+G results. The agreement is excellent, although the small variation of the $3s$ and $3p$ branching ratios as a function of the collision energy is not present in the theoretical calculations.
The theoretical branching ratios obtained with the different basis sets discussed in Section \ref{sec:theory} are given in Table~\ref{table_BR} where they can be compared to the experimental values as well as with the theoretical calculations of \citet{croft_rate_1999}. Results obtained with the ACV5Z, ET-mid, and ET-Li basis sets are also shown in Fig. \ref{fig_BR}.  The theoretical values given in table \ref{table_BR} are valid for all energies below 0.5 eV, as the branching ratio is predicted to be energy-independent in this energy range.
The branching ratios of \citet{croft_rate_1999} are in good agreement with our ACV5Z+G results for the $3s$ and $3p$ states, while the contribution of the $3d$ state is too low by a factor of three. 

An important observation concerns the performance of the other basis sets used in this work. 
We see from table \ref{table_BR} that the branching ratios calculated based on the results obtained with the ACV5Z and ET-Li basis sets both show a very small contribution from the $3s$ state, in complete disagreement with the experimental results. The branching ratios obtained with the ET-mid basis set seem to be in qualitative agreement with the experimental data. However, a contribution of 25\% of the $n=2$ states is predicted, which again disagrees with the experimental data. 
These results illustrate the strong dependence of reaction cross sections towards \textit{ab initio} data, particularly in cases for which many avoided crossings contribute, as is the case for the mutual neutralization processes. 

The experimental branching ratios were also obtained for Li$^+$ + H$^-$ collisions at 3 meV.  
The MN branching ratios were 64.1\% for the $3s$ state, 28.0\% for the $3p$ state, and 7.9\% for the $3d$ state. By comparing with the results presented Table \ref{table_BR}, these values are seen to be similar to those measured for Li$^+$-D$^-$, and we observed no clear isotope effect.

\begin{table*}
\begin{center}
\begin{tabular}{l|ccccccc}
			& Li($2s$) & Li($2p$) & Li($3s$) 	& Li($3p$) 	& Li($3d$) & Li($4s$)	  	\\ \hline
Expt. 3.9 meV	& -- &--& 63.4(2.1) 		& 26.6(1.2) 		& 10.0(0.7) &	--	\\
Expt. 20 meV	& -- & -- & 60.0(2.0) 		& 32.0(1.0) 		& 8.0(1.0) &	--	\\
Expt. 200 meV	& -- & -- & 59.9(4.0) 		& 33.7(7.0) 		& 6.5(2.0) &	--	\\
ACV5Z+G		& 0 & 0 & 60.1			& 30.5			& 8.9 	&	0.5	\\
ACV5Z		& 0 & 0 & 15.5			& 75.0			& 9.1 	&	0.4	\\
ET-mid	  	& 0.3 & 24.4 & 52.2			& 17.5			& 5.1 	&	0.5	\\
ET-Li 		& 0 & 0 & 11.8			& 42.0			& 45.8 	&	0.3	\\
\cite{croft_rate_1999} & -- & -- & 66	 	& 31				& 3	 &	--	\\
\hline \hline
\end{tabular} 
\end{center}
\caption{Experimental branching ratios (in \%) obtained at three energies (3.9 meV, 20 meV, and 200 meV) for the mutual neutralisation Li$^+$ + D$^-$ $\rightarrow$ Li($nl$) + D, compared to the theoretical results obtained with various basis sets and with the calculations of \citet{croft_rate_1999}. The uncertainty is given in parentheses. The theoretical branching ratios are energy-independent in this energy range.}
\label{table_BR}
\end{table*}

\subsection{Rate coefficients}

The study of lithium abundances in stellar atmospheres provides important information about stellar evolution and Big-Bang nucleosynthesis \citep{Barklem_2003_LiH}. The analysis of the Li I 670.8 nm and 610.4 nm lines through non-LTE models \citep{Asplund_2005,barklem_accurate_2016}, allowing the determination of relative and absolute abundances, requires accurate rate coefficients for the mutual neutralization reaction. 
In this context, updated values for this process as well as for inelastic Li--H collisions were recently presented by \citet{Belyaev_2018_LiH}. These data rely on the \textit{ab initio} results of \citet{croft_theoretical_1999}.

Based on the cross sections measured and calculated in the present work, we can 
compute mutual neutralization rate coefficients for both Li$^+$--H$^-$ and Li$^+$--D$^-$ collisions. In Table \ref{table_rates_comp} we compare our results for Li$^+$--H$^-$ and Li$^+$--D$^-$ with those obtained by \citet{Belyaev_2018_LiH} for Li$^+$--H$^-$ collisions at three temperatures (2000K, 6000K, and 10000K). Over this range of temperatures, the total rate coefficient for Li$^+$--H$^-$ is larger than for Li$^+$--D$^-$ by about 25\%. 
There is a small isotope dependency of the branching ratios. At 10000~K, for Li$^+$--D$^-$ the $3s$, $3p$, and $3d$ states contribute respectively to 63\%, 28\%, and 8\% of the total rate coefficient, while for Li$^+$--H$^-$ we find values of 68\%, 24\%, and 7\%. 
Our calculated total rate coefficient for Li$^+$--H$^-$ is smaller than the result of \cite{Belyaev_2018_LiH} by about 15\% at 2000~K and 25\% at 10000~K. Moreover, the branching ratios of \cite{Belyaev_2018_LiH} are 69\%, 28\%, and 2.6\% for the $3s$, $3p$, and $3d$ states, respectively. As a result, the absolute partial rate coefficients for the $3s$ and $3p$ states are smaller by about 20\% and 40\%, respectively, than the results presented by \cite{Belyaev_2018_LiH}, while for the $3d$ state our results are larger by a factor of 2.

\begin{table*}
\hspace{-4cm}
\begin{tabular}{c|ccc|ccc|ccc|ccc}
& \multicolumn{3}{c}{Li($3s$)}	& \multicolumn{3}{c}{Li($3p$)}	& \multicolumn{3}{c}{Li($3d$)} & \multicolumn{3}{c}{Total} \\ 
$T$(K)		& D	& H	& H\tablenotemark{(a)}  & D	& H	& H\tablenotemark{(a)} & D	& H	& H\tablenotemark{(a)}	& D	& H	& H\tablenotemark{(a)} \\ \hline
2000		& 7.54[-8] 	& 1.04[-7]	& 1.21[-7]	& 3.72[-8]	& 3.96[-8]	& 5.36[-8]	& 1.09[-8]	& 1.13[-8]	& 4.92[-9]	& 1.24[-7]	& 1.56[-7]	& 1.80[-7]	\\
6000		& 5.54[-8]	& 7.43[-8]	& 9.47[-8]	& 2.57[-8]	& 2.72[-8]	& 4.00[-8]	& 7.48[-9]	& 7.72[-9]	& 3.71[-9]	& 8.90[-8]	& 1.10[-7] 	& 1.39[-7]	\\
10000	& 5.15[-8]	& 6.76[-8]	& 8.49[-8]	& 2.27[-8]	& 2.39[-8]	& 3.48[-8]	& 6.57[-9]	& 6.77[-9]	& 3.19[-9]	& 8.12[-8]	& 9.87[-8] 	& 1.23[-7]	\\
\hline \hline
\end{tabular} 
\tablenotetext{a}{  From \citet{Belyaev_2018_LiH}}
\caption{Rate coefficients for Li$^+$ + H$^-$/D$^-$ $\rightarrow$ Li($nl)$ + H/D at three temperatures, compared with the theoretical results of \citet{Belyaev_2018_LiH} for Li$^+$ + H$^-$. Brackets indicate powers of ten.}
\label{table_rates_comp}
\end{table*}

In order to facilitate their use in non-LTE models, the rate coefficients for Li$^+$ + H$^-$/D$^-$ $\rightarrow$ Li($nl)$ + H/D are fitted to the modified Arrhenius equation, 
\begin{equation}\label{eq_fit_rate}
k(T)=\alpha \left( T/300 \right)^\beta e^{-\gamma/T} \ .
\end{equation}
The fitting parameters are given in table \ref{table_rates_fit} for the total rate coefficient as well as for the partial rate coefficients for the Li($n=3$) states. The accuracy of the fit is estimated to be 5\% in the range from 500K to 10000K.

\begin{table*}
\begin{center}
\begin{tabular}{l|cc|cc|cc}
		& \multicolumn{2}{c}{$\alpha$(cm$^3$\, s$^{-1}$)}	& \multicolumn{2}{c}{$\beta$}		& \multicolumn{2}{c}{$\gamma$ (K)}  	\\ 
		& H	& D	& H	& D	& H	& D \\ \hline
Li($3s$)	& 1.478$\times 10^{-7}$	& 8.117$\times 10^{-8}$	& $-0.238$	& $-0.130$	& $-205.6$ 	& $-289.4$ 	\\
Li($3p$)	& 6.305$\times 10^{-8}$	& 5.263$\times 10^{-8}$	& $-0.288$	& $-0.240$	& $-159.0$	& $-187.6$	\\
Li($3d$)	& 1.824$\times 10^{-8}$	& 1.582$\times 10^{-8}$	& $-0.294$	& $-0.251$	& $-153.6$	& $-177.4$	\\
Total		& 2.295$\times 10^{-7}$	& 1.483$\times 10^{-7}$	& $-0.254$	& $-0.172$	& $-190.3$	& $-250.6$	\\
\hline \hline
\end{tabular} 
\end{center}
\caption{Parameters of the fit of the rate coefficients (Eq. (\ref{eq_fit_rate})) for Li$^+$ + H$^-$/D$^-$ $\rightarrow$ Li($nl$) + H/D mutual neutralization. }
\label{table_rates_fit}
\end{table*}

\section{\label{sec:conclusion}Conclusion}

Using a merged-beam apparatus, we have studied the mutual neutralization process in collisions of Li$^+$ with D$^-$ at energies from 1 eV down to the meV range. In addition to measurements of the total cross section, we were able to analyze the kinetic energy release, giving access to the state-to-state cross sections and branching ratios. We showed that following electron capture the only electronic states of the Li($1s^2\ nl$) products that are significantly populated are the $n=3$ states.
We also investigated the mutual neutralization process theoretically by performing multichannel Landau-Zener calculations that rely on \textit{ab initio} quantum chemistry calculations. We obtained the potential energy curves and non-adiabatic couplings of the LiH molecule with several basis sets. The calculated total mutual neutralization cross section is in good agreement with the present measurements and with previous theoretical calculations. On the other hand, for the branching ratios we found discrepancies between our measurements and previous calculations, particularly for the Li($1s^2\ 3d$) state. Moreover, we showed that the branching ratios are extremely sensitive to the basis set employed in the \textit{ab initio} calculations. This demonstrates the importance of the accuracy of the \textit{ab initio} quantum chemistry methods in order to study elementary reactive processes involving several excited electronic states as well as the crucial role of low-energy experiments to validate the widely-employed theoretical tools. 

The total and partial cross sections obtained in this work were used to revisit the total and partial rate coefficients for the mutual neutralization reaction in Li$^+$--H$^-$ and Li$^+$--D$^-$ collisions, and we provided new reference values for this process over a wide range of temperatures that can be used in astrochemical models.

\acknowledgments
T. Launoy would like to thank P. Quinet and P. Palm\'eri from the Umons university for the {\small AUTOSTRUCTURE} results.

This work was supported by the I.I.S.N. instrument (grant 4.4504.10) of the Fonds de la Recherche Scientifique - FNRS.
Computational resources have been provided by the Shared ICT Services Centre of the Universit\'e libre de Bruxelles 
and by the Consortium des \'Equipements de Calcul Intensif (C\'ECI), funded by the Fonds de la Recherche Scientifique de Belgique (F.R.S.-FNRS) under Grant No. 2.5020.11.
The authors thank the Belgian State for the grant allocated by Royal Decree for research in the domain of controlled thermonuclear fusion. 
XU is Senior Research Associate of the Fonds de la Recherche Scientifique - FNRS.

\end{document}